\documentclass{article}
\usepackage{authblk}
\usepackage{amsmath,amsfonts}

\begin{document}

\title{The fundamental connections between classical Hamiltonian mechanics, quantum mechanics and information entropy}

\author[\dagger]{Gabriele Carcassi}

\author[\dagger]{Christine A. Aidala}

\affil[\dagger]{Physics Department, University of Michigan, 450 Church Street\\
Ann Arbor, MI 48109-1040,
United States\\
carcassi@umich.edu}

\maketitle

\begin{abstract}
We show that the main difference between classical and quantum systems can be understood in terms of information entropy. Classical systems can be considered the ones where the internal dynamics can be known with arbitrary precision while quantum systems can be considered the ones where the internal dynamics cannot be accessed at all. As information entropy can be used to characterize how much the state of the whole system identifies the state of its parts, classical systems can have arbitrarily small information entropy while quantum systems cannot. This provides insights that allow us to understand the analogies and differences between the two theories.
\end{abstract}



\markboth{Carcassi, Aidala}
{The Fundamental Connections between Mechanics and Information Entropy}

\section{Introduction}
In a previous work\cite{Carc1} we identified a small set of physical assumptions from which classical and quantum particle mechanics can be rederived. By assumptions here we mean simplifying conditions that our system is supposed to satisfy under the processes under study. Our work differs from others, see for example Refs.~\cite{PhysRevA.84.012311,QLogicReview,Hardy:2001jk,ludwig_hein_2013}, in that the starting points are physical assumptions about the system being studied (e.g. whether the parts can be studied or whether the evolution is deterministic) and not information theoretic or abstract mathematical considerations. It is also different in that both classical and quantum cases are derived on equal footing. That is, we have identified a single point where one assumption leads to the classical case while another leads to the quantum system.

The difference is the answer to a single question: is the internal dynamics of the system accessible or not? If it is fully accessible we recover classical states and if it is totally inaccessible we recover quantum states. This difference can be understood in terms of information entropy and in this work we develop that insight. We will present here a much shorter version of the classical derivation that carries all the important points. For quantum mechanics, for the sake of brevity, we will cover only some of the general ideas. For details on both, we refer to the previous longer paper\cite{Carc1}. We then provide a classical analogue of the quantum uncertainty relationship and see how both can be understood in terms of information entropy conservation under the different initial assumptions.

We believe these insights better clarify the difference between classical and quantum systems, their limit of validity and therefore where one can expect them to fail.

\section{Reducible Systems and Classical Phase Space}
We define a classical system as one that is assumed to be infinitesimally reducible. That is, the system can be thought as made of indefinitely small parts and giving the state of the system is equivalent to giving the state of its parts. For example, we can throw a ball and study its motion; or we can take a red marker, make a dot on the ball and study the motion of the red dot. Studying the motion of the whole ball is equivalent to studying the motion of all the possible red dots we could draw. To a first approximation, a ball can be thought as infinitesimally reducible, made of a continuous material.

If we assume our system is reducible, then the state of the whole system is simply a distribution over the states of the infinitesimal parts, which we call particles. That is, our state is a normalized integrable function
\begin{equation}
\begin{aligned}
	\rho : \mathcal{S} \to \mathbb{R} \\
	\int_{\mathcal{S}} \rho ds = 1
\end{aligned}
\end{equation}
over the particle state space $\mathcal{S}$. If we have a set of state variables $\xi^a : \mathcal{S} \to \mathbb{R}$ that are sufficient to characterize our particles, we can write both the density and the information entropy
\begin{equation}
\begin{aligned}
\rho(\xi^a) &= \rho(s(\xi^a)) \\
I[\rho] &= - \int_{\mathcal{S}} \rho \log (\rho) d\xi^1 d\xi^2 ...
\end{aligned}
\end{equation}
in terms of the state variables.

The issue is that changing coordinate system will change the numeric value of the density. Yet, the density is fully specified by the particle states and, since states are invariant under coordinate transformations, it should be invariant. That is, we have
\begin{equation}
\begin{aligned}
\rho(\hat{\xi}^b) &= \left| \frac{\partial \hat{\xi}^b}{\partial \xi^a} \right| \rho(\xi^a) \\
\rho(s(\hat{\xi}^b)) &= \rho(s(\xi^a)) \\
\end{aligned}
\end{equation}
which means our distribution $\rho$ has to both transform as a scalar (i.e. it fully depends on the state) and as a density (i.e. the integral needs to be invariant).

This apparent contradiction is actually a formal requirement that severely limits what types of manifolds can be used to describe particle states. It turns out that these are precisely the symplectic manifolds: the structure of phase space is exactly the one required to write coordinate invariant distributions. That is, given a set of coordinates $q^i$ (state variables that define the reference frame) we will need another set of state variables $k_i$ with inverse units such that their product will become an invariant pure number. Under a general coordinate transformation $\hat{q}^j = \hat{q}^j (q^i)$ we will simply have 
\begin{equation}
\begin{aligned}
dq^i dk_i = d\hat{q}^j d\hat{k}_j
\end{aligned}
\end{equation}
which means the Jacobian of the transformation is unitary. The invariance of areas is equivalent to the invariance of densities and of information entropy. In other words, the structure of phase space is precisely the one that allows us to define information entropy as coordinate independent. Only on these spaces information entropy can be a physically well-defined quantity over continuous variables.

If we require deterministic and reversible evolution, we require that the density over one state is transported exactly to another state. Equivalently, we require that the information given at one time to identify a part is exactly the same to identify the same part at a different time: information entropy is the same. In this case we are requiring that the time evolution is an area preserving (i.e. canonical) transformation. This gives us Hamilton's equations. This tells us that the following four concepts are very tightly related:
\begin{itemize}
	\item Deterministic and reversible evolution - state densities are mapped one-to-one
	\item System isolation - the evolution of the system depends only on its state and on nothing else
	\item Conservation of information entropy - the information required to describe the system does not change in time
	\item Conservation of energy - Hamiltonian evolution.
\end{itemize}

\section{Irreducible systems and quantum mechanics}

Conversely, we define a quantum system as one that is assumed to be irreducible. That is, the system can still be thought as made of parts but the state of the system tells us nothing about the state of the parts. For example, we can isolate an electron and study its motion; yet we cannot put a red mark on part of the electron, we cannot scatter a photon off of only a part of an electron. As the interaction is always with the whole electron, no information can be gained about its internal dynamics. An electron can be thought as irreducible.

We stress that both these assumptions are operational and depend not only on the system, but also on the processes we are studying or available to us. For example, a proton can be modeled as a single quantum system and will behave as such provided the internal dynamics is not relevant. It will exhibit diffraction and interference patterns, have half-integer spin and so on. If we interact in a way that disturbs the internal dynamics, however, we will see that there is internal dynamics in terms of quarks and gluon. In those regimes the proton cannot be treated as a single quantum system, but must be treated as a composite system.

To summarize the arguments laid out in \cite{Carc1}, since the state of the system does not capture the state of the parts, we can imagine these continuously permutating without affecting the dynamics of the whole. The internal motion is then characterized by random variables, which still need to allow for the definition of an invariant density, and therefore they must, for a degree of freedom, form a pair $(U,V)$. As the value of the variable is completely unknown\footnote{One may be tempted to distinguish between the following cases: the internal dynamics is unknown, the internal dynamics is unknowable, the internal dynamics is absent/not defined. In fact, different interpretations of quantum mechanics may take different positions. However, one has to be clear how those cases are going to differ experimentally, or this distinction remains within the philosophical realm. In our case, we are limiting ourselves to posit that the processes we are studying are not sensitive to a further description of the system, which makes a possible internal dynamics automatically unknown. It could be that we will find new processes in the future, it could be that such processes are forever outside of what we can control experimentally or it could be that they do not even exist. We cannot know which case we are in: all we is that know we are not sensitive. As we do not have a way to tell, these cases are not distinguishable to us experimentally and they can be treated as the same case.}, the distribution will be uniform.\footnote{The addition of extra degrees of freedom that do not alter the problem but make it easier to study is something common in math and physics. It is common in probability theory, where adding random variables is a standard technique in proofs. It is common in analysis when extending a real integral to the complex plane. It is common in statistical mechanics in order to study system/environment interactions. It is common in quantum mechanics with purification. In all these cases, we are extending our space while making sure that the result does not depend on the particulars of how it was extended. Adding an internal dynamics to which we are insensitive is just another instance.} We can fix the density $\rho=1$ to be unitary and the standard deviations $\sigma_U=\sigma_V$ to be equal as this internal relabelling cannot affect the state of the system as a whole. The transformations that will affect the system as a whole are the changes in variance and correlation as these would change the value of the integral $\int \rho dudv \propto \sigma_U\sigma_V = \sigma_U^2$. These are the transformations of the type:
\begin{equation}
	\begin{aligned}
	\hat{U} = a U + b V \\
	\hat{V} = -b U + a V
	\end{aligned}
\end{equation}
A complex number $c=a + \imath b$ can be used to identify such transformations. The norm represents the strength of the internal random process, which is proportional to the size of the system, and the phase relates to the correlation as the arccosine of the Pearson coefficient.

We want to stress, again, that we consider the classical state to be a matter distribution $\rho(q^i, k_i)$. Therefore, when comparing classical and quantum mechanics, we should be comparing the properties of our wave function $\psi(q^i)$ to the matter distribution $\rho(q^i, k_i)$, not to points of phase space. Comparing the two mechanics, in fact, means comparing how the same system would be different under the different assumptions. So we have to compare finite systems to finite systems. In both cases, the ``particle'' is the smallest part of the system to which we can assign a state. Yet, in quantum mechanics it is itself the indivisible system, while a classical particle is not a system: it is the limit of recursive subdivision.  While this different comparison may at first seem awkward, as it is not the usual framing, it leads to a more stringent parallel between the mathematical elements of classical and quantum mechanics and, in turn, to a better intuition. 

For example, it makes it clear that the position of classical objects (e.g. planets, cannonballs, drops of water and even ourselves) cannot be fully described by a single number, as it is often said. All these objects have extent both in space and momentum. A point-particle is the limit where the extent of the distribution can be considered small, and therefore the center of mass is enough for our description. This limit can be performed within both classical and quantum mechanics if the scale of the problem allows, and, because of Ehrenfest's theorem, the behavior is the same. In light of the assumptions, it should be obvious this should be the case: the difference between classical and quantum is in the description of the parts; if the length scales of interest are much greater than those of the system, then it does not matter what we assume about the parts.

\section{Uncertainty principles and information entropy}

Since classical Hamiltonian evolution conserves information entropy, the space of distributions explored by time evolution is limited by that constraint. We then ask the following question: what is the distribution that, given a certain amount of information entropy, minimizes the spread over phase space? That is, what is $\rho(q, k)$ such that $\sigma_q\sigma_k$ is minimized? Given the constraint, no Hamiltonian evolution can lead to a state less spread than that.

It turns out this distribution is the product of two Gaussian distributions
\begin{equation}
\rho(q,k) = \frac{1}{2\pi\sigma_q\sigma_k} e^{- \frac{(q - \mu_q)^2}{2 \sigma_q^2}} e^{- \frac{(p - \mu_p)^2}{2 \sigma_p^2}}
\end{equation}
with information entropy
\begin{equation}
I_G = \ln (2 \pi e \sigma_q \sigma_k).
\end{equation}
Since this is the distribution that minimizes the spread, during the evolution we have
\begin{equation}
\sigma_q \sigma_k \leq \frac{e^{I_0}}{2 \pi e}
\end{equation}
where $I_0$ is the initial information entropy. This classical uncertainty relationship is reminiscent of the quantum uncertainty relationship in both the form and the fact that the equality holds for the Gaussian packets. The analogy can be made stronger with a few extra conditions as described in Ref.~\cite{Carc1}. What we want to show here is the connection between this uncertainty relationship, information entropy, deterministic and reversible evolution and the difference in reducibility between a classical and a quantum system.

First of all we note that this classical uncertainty relationship is given by conservation of entropy and therefore deterministic and reversible evolution. We are free to set the initial amount of information entropy for the classical distribution as small as we want. Once it is set, this cannot be changed under Hamiltonian evolution. Quantum mechanics has a stronger condition: we cannot set the initial uncertainty as small as we want. It is bound by a fundamental constant. For quantum states, all pure states have the same information entropy which is set to zero. This difference is a direct consequence of the difference in reducibility: in classical mechanics we can always access the internal dynamics, and therefore we can create narrower and narrower distributions (e.g. we can always divide a continuous material) and study their individual evolution; in quantum mechanics we cannot (e.g. we cannot prepare half an electron) so all states necessarily are associated with the same information entropy. The entropy for a pure state, then, is set to zero to signify that the dynamics within a quantum system is not accessible, there is no information to be had.

The deterministic and reversible case in quantum mechanics is given by unitary evolution, which in fact conserves von Neumann entropy because each pure state is mapped to a pure state. Moreover, given two pure quantum states of the same system we can always find a unitary evolution that connects the two (at least mathematically) while this is not possible in classical Hamiltonian mechanics (e.g. two distributions with different entropy cannot be linked by Hamiltonian evolution).

These insights lead us to conclude that the difference between a classical and quantum system is not primarily about its size. The previous discussion does not depend on the physical extent of the system or the number of its degrees of freedom. It is about the accessibility of the internal dynamics. If we are to look at scenarios where classical mechanics fails, we simply have to look at cases where we can no longer assume that the internal dynamics is indefinitely accessible. Conversely, if we were to look at scenarios where quantum mechanics fails as applied to a specific system, we would need to look at cases where the internal dynamics of that system becomes at least minimally accessible.

\section{Conclusion}

We have shown that the principal difference between classical and quantum systems is how much the state of the whole can characterize its parts. In classical systems the internal dynamics is fully accessible while in a quantum system it is completely inaccessible. This can be understood in terms of information entropy: classical states as distributions over phase space allow for arbitrarily small amounts of information entropy while pure quantum states have a fixed entropy which is set to zero.

\section*{Acknowledgments}

Funding for this work was provided in part by the MCubed program of the University of Michigan.

\bibliographystyle{plain}
\bibliography{bibliography}

\begin{thebibliography}{1}

\bibitem{Carc1}
Gabriele Carcassi, Christine~A. Aidala, David~J. Baker, and Lydia Bieri.
\newblock From physical assumptions to classical and quantum {Hamiltonian} and
  {Lagrangian} particle mechanics.
\newblock {\em Journal of Physics Communications}, 2(4):045026, apr 2018.

\bibitem{PhysRevA.84.012311}
Giulio Chiribella, Giacomo~Mauro D'Ariano, and Paolo Perinotti.
\newblock Informational derivation of quantum theory.
\newblock {\em Phys. Rev. A}, 84:012311, Jul 2011.

\bibitem{QLogicReview}
Bob Coecke, David Moore, and Alexander Wilce.
\newblock {\em Current Research in Operational Quantum Logic: Algebras,
  Categories, Languages}.
\newblock 01 2000.

\bibitem{Hardy:2001jk}
Lucien Hardy.
\newblock {Quantum theory from five reasonable axioms}.
\newblock 2001.
\newblock quant-ph/0101012.

\bibitem{ludwig_hein_2013}
G.~Ludwig and C.~A. Hein.
\newblock {\em Foundations of Quantum Mechanics}.
\newblock Springer Berlin, 2013.

\end{thebibliography}

\end{document}